\documentclass[twocolumn,numberedappendix,iop]{emulateapj}
\shorttitle{Modeling Collisional Cascades: Steep Dust-Size Distributions}
\shortauthors{G\'asp\'ar et al.}
\journalinfo{The Astrophysical Journal, Accepted: May 21, 2012}
\usepackage{amsmath}
\usepackage{multirow} 

\begin{document}

\title{Modeling Collisional Cascades In Debris Disks: Steep Dust-Size Distributions}

\author{Andr\'as G\'asp\'ar}
\author{Dimitrios Psaltis}
\author{George H.\ Rieke}
\author{Feryal \"Ozel}
\affil{Steward Observatory, University of Arizona, Tucson, AZ 85721\\
agaspar@as.arizona.edu, dpsaltis@as.arizona.edu, grieke@as.arizona.edu, fozel@as.arizona.edu}

\begin{abstract}
We explore the evolution of the mass distribution of dust in collision-dominated debris 
disks, using the collisional code introduced in our previous paper. We analyze the 
equilibrium distribution and its dependence on model parameters by evolving over 100 
models to 10 Gyr. With our numerical models, we confirm that systems reach collisional 
equilibrium with a mass distribution that is steeper than the traditional solution by \cite{dohnanyi69}. 
Our model yields a quasi steady-state slope of 
$n(m) \sim m^{-1.88}$ [$n(a) \sim a^{-3.65}$] as a robust solution for a wide range of
possible model parameters. We also show that a simple
power-law function can be an appropriate approximation for the mass distribution
of particles in certain regimes. The steeper solution has observable 
effects in the submillimeter and millimeter wavelength regimes of the electromagnetic 
spectrum. We assemble data for nine debris disks that have been observed at these 
wavelengths and, using a simplified absorption efficiency model, show that the predicted 
slope of the particle mass distribution generates SEDs that are in agreement with the 
observed ones.
\end{abstract}
\keywords{methods: numerical -- circumstellar matter -- planetary systems -- infrared: stars}

\section{Introduction}

The total mass within debris disks as well as the infrared excess emission produced by 
their dust are generally calculated assuming the analytic estimate of the distribution of 
masses in the asteroid belt by \cite{dohnanyi69}. This solution predicts that the sizes 
follow a power-law, with their numbers increasing with decreasing size $a$ as 
$n(a) \sim a^{-3.5}$. However, a number of recent efforts  to model observations of 
debris disks have found it necessary to adopt steeper slopes \citep{krist10,golimowski11}.

\cite{durda97} and \cite{obrien03} have shown numerically and analytically
that a steep tensile strength curve, i.e., the function that gives the minimum 
energy required to disrupt a body catastrophically \citep[see, e.g.,][]{holsapple02,benz99,gaspar12a}, 
results in a steeper quasi steady-state distribution than the traditional solution. Collisional models 
of the dust in debris disks \citep{thebault03,krivov05,thebault07,lohne08,muller10,wyatt11} 
have also shown that the dust particles will settle with a size distribution
slope larger than 3.6, on top of which additional structures appear. This steeper distribution has readily observable 
effects at the far-IR and submm wavelengths. It also results in higher total dust mass and 
lower planetesimal mass estimates for the systems. However, these results are 
often disregarded in observational studies. Instead, the traditional \cite{dohnanyi69} 
slope of mass distribution is used, likely due to the complexity of the numerical models, the superposed
wavy structures on the distributions, and their uncertain regimes of applicability.

In this paper, we investigate the slope of the mass distribution and the physical parameters 
that influence it with the numerical code introduced in \citeauthor{gaspar12a} (2012, hereafter 
Paper I). Our code calculates the evolution of the particle mass distribution in 
collisional systems, taking into account both erosive and catastrophic collisions. 
In \S \ref{models}, we introduce models for the numerical analysis of the collisional cascades and give our findings. 
In \S \ref{sec:approx}, we analyze the ranges where a simple power-law function is an 
appropriate approximation for the mass distribution of particles.
In \S \ref{sec:companal}, we compare our model's properties to the assumptions 
of previous analytic solutions, while in
\S \ref{sec:synthetic}, we generate a set of synthetic spectra in order to analyze the effects 
certain distribution parameters have on different parts of the SEDs. In \S \ref{sec:relation}, 
we introduce a simple relation between the Rayleigh-Jeans part of the spectral energy distributions 
and the particle size distribution. In \S \ref{sec:constraints}, we compare our results to the observed 
far-IR and sub millimeter data for nine sources.

\section{Numerical modeling}
\label{models}

In this section, we analyze the dust distribution with our full numerical code (Paper I).
We run a set of numerical models to study the evolution of the slope of the distribution
function and its dependence on the model parameters. We investigate the time required for the 
distribution to settle into its quasi steady-state and, with a wide coverage of the parameter space, we 
also examine the robustness of the solution.

\subsection{Evolution of the reference model}
\label{reference}

\begin{deluxetable*}{llr}
\tablecolumns{3}
\tabletypesize{\scriptsize}
\singlespace
\tablewidth{0pt}
\tablecaption{Numerical, Collisional, and System parameters of our reference model\label{tab:tabvar}}
\tablehead{
\colhead{Variable} & \colhead{Description} & \colhead{Fiducial value} }
\startdata
\multicolumn{3}{c}{System variables}					 														\\
\hline\hline																								
$\rho$		 & Bulk density of particles \dotfill											& 2.7 g cm$^{-3}$			\\
$m_{\rm min}$ 	 & Mass of the smallest particles in the system \dotfill							& 1.42$\times10^{-21}$ kg	\\
$m_{\rm max}$ & Mass of the largest particles in the system \dotfill								& 1.13$\times10^{22}$ kg		\\
$M_{\rm tot}$	 & Total mass within the debris ring \dotfill									& 1 $M_{\Earth}$ 			\\
$\eta_0$	 	 & Initial power-law distribution of particle masses \dotfill							& 1.87 					\\
$R$		 	 & Distance of the debris ring from the star \dotfill 								& 25 AU 					\\
$\Delta R$	 & Width of the debris ring \dotfill											& 2.5 AU 					\\
$h$		 	 & Height of the debris ring \dotfill											& 2.5 AU 					\\
Sp	 		 & Spectral-type of the star \dotfill											& A0 					\\
\hline																									
\multicolumn{3}{c}{Collisional variables}																			\\
\hline\hline																								
$\gamma$	 & Redistribution power-law \dotfill											& 11/6 					\\
$\beta_X$	 	 & Power exponent in X particle equation \dotfill								& 1.24 					\\
$\alpha$		 & Scaling constant in $M_{\rm cr}$ \dotfill									& $2.7\times10^{-6}$ 		\\
$b$			 & Power-law exponent in $M_{\rm cr}$ equation \dotfill							& 1.23 					\\
$f_M$		 & Interpolation boundary for erosive collisions \dotfill							& $10^{-4}$ 				\\
$f_Y$		 & Fraction of $Y/M_{\rm cr}$ \dotfill											& 0.2 					\\
$f_X^{\rm max}$  & Largest fraction of $Y/X$ at super catastrophic collision boundary \dotfill 			& 0.5 					\\
$Q_{\rm sc}$	 & Total scaling of the $Q^{\ast}$ strength curve \dotfill							& 1 						\\
$S$			 & Scaling of the strength regime of the $Q^{\ast}$ strength curve \dotfill				& $3.5\times10^{7}$ erg/g 	\\
$G$			 & Scaling of the gravity regime of the $Q^{\ast}$ strength curve \dotfill				& 0.3 erg cm$^3$/g$^2$		\\
$s$			 & Power exponent of the strength regime of the $Q^{\ast}$ strength curve \dotfill		& -0.38 					\\
$g$			 & Power exponent of the gravity regime of the $Q^{\ast}$ strength curve \dotfill 		& 1.36  					\\
\hline																									
\multicolumn{3}{c}{Numerical parameters}																		\\
\hline\hline																								
$\delta$	 	& Neighboring grid point mass ratio \dotfill									& 1.104			 		\\
$\Theta$	 	& Constant in smoothing weight for large-mass collisional probability\dotfill			& $10^6 m_{\rm max}$		\\
$P$		 	& Exponent in smoothing weight for large-mass collisional probability\dotfill			& 16				
\enddata
\end{deluxetable*}

We set up a reference debris disk as a basis for comparison to all other model runs. This model 
consists of a moderately dense debris disk situated at 25 AU around an A0 spectral-type star, with 
a width and height of 2.5 AU. This radial distance ensures a moderate evolution speed. The 
peak emission is in the mid-infrared, with a Rayleigh-Jeans tail in the far-infrared regime, 
which is the primary imaging window for the {\it Herschel Space Telescope}. The total mass in the 
debris disk is $1~M_{\Earth}$, distributed between minimum and maximum particle masses that 
correspond to radii of  5 nm and 1000 km, when assuming a bulk density of 2.7 g cm$^{-3}$. 
We summarize the disk parameters of the reference model in Table \ref{tab:tabvar}. We evolve the 
reference model for $10~{\rm Gyr}$.

\begin{figure}[!ht]
\begin{center}
\includegraphics[angle=0,scale=0.7]{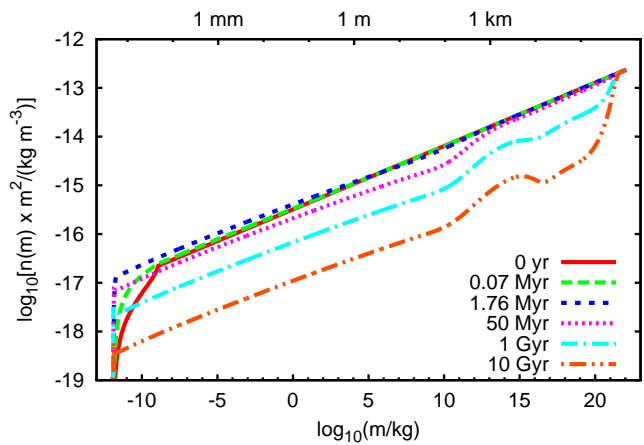}
\caption{Particle mass distribution of the reference model plotted at various points in time.}
\label{fig:evolve}
\end{center}
\end{figure}

In Figure \ref{fig:evolve}, we show the evolution of the particle distribution, plotting it at six different 
points in time up to 10 Gyr. On the vertical axis we plot $\log_{10}\left[n\left(m\right)\times m^2\right]$, 
which can be related to the ``mass/bin'' that is frequently used in other simulations. Even though the 
number densities decrease with increasing particle masses, the mass distribution increases towards 
the larger masses in this representation, as long as the mass distribution slope is smaller than 2.

The smallest particles reach collisional equilibrium first, roughly at $1~{\rm Myr}$, followed by larger 
particle sizes as the system evolves. After $50$-$100~{\rm Myr}$ of evolution, the upper, gravity 
dominated part of the distribution ($m > 10^{13}~{\rm kg}$) also reaches equilibrium. The distribution 
maintains its slope for masses below $10^{10}~{\rm kg}$, which roughly corresponds to a planetesimal 
radius of $100~{\rm m}$. The kink in the distribution at the upper end is due to the change in the strength 
curve slope \citep{obrien05,bottke05}.

Structures in the distribution slope, such as waves, may in principle occur at the low-mass end when 
assuming softer material properties or higher collision velocities (see Section \ref{sec:approx}). 
The distribution may also acquire some 
curvature (see Paper I). Because of these effects, we evaluate the average slope of the distribution
by fitting a power-law over a large mass range, but one that remains below the kink in the distribution. 
Specifically, we fit the slope between masses $10^{-8}$ and $10^4$ kg, which roughly correspond 
to sizes of 0.1 mm and 1 m.

\begin{figure*}[!ht]
\begin{center}
\includegraphics[angle=0,scale=1.4]{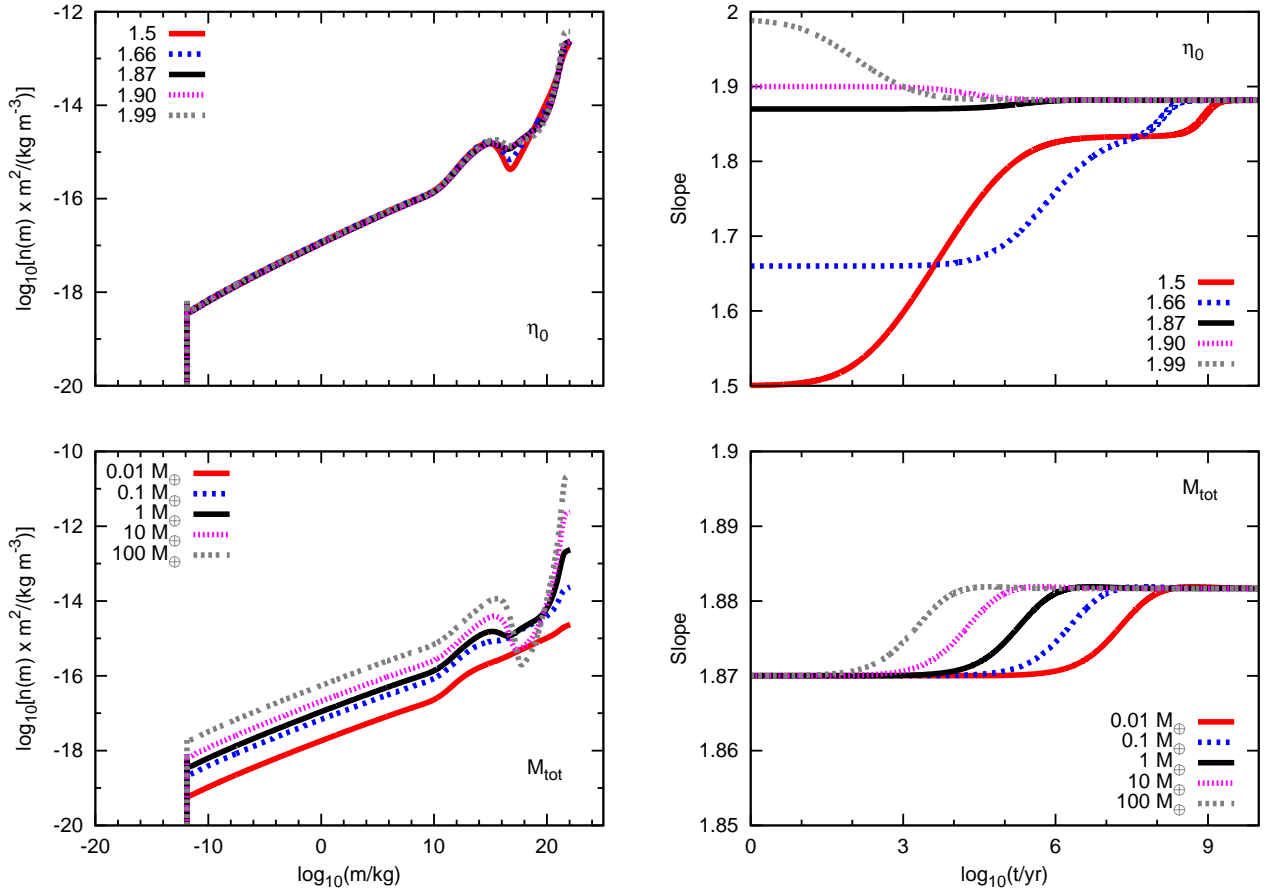}
\caption{({\it Left panels}) Particle mass distribution at $10~{\rm Gyr}$, when
varying the initial mass distribution slope (top) and the total mass of the system(bottom). ({\it Right panels}) Evolution
of the dust-mass distribution slopes when varying the initial mass distribution slope (top) and the total mass of the system (bottom).
The quasi steady-state distribution slope is practically independent of these initial conditions.}
\label{fig:etaM}
\end{center}
\end{figure*}

We next examine the dependence of the quasi steady-state distribution slope on the initial conditions , i.e.~
the initial mass-distribution slope $\eta_0$ and the initial total mass in the disk 
$M_{\rm tot}$. Figure \ref{fig:etaM} shows the evolution of the particle mass distribution and 
its slope as a function of these initial conditions. The left panels show the mass distribution after $10~{\rm Gyr}$ 
for different values of the two input parameters, while the right panels show the evolution of 
the dust-mass distribution slope. Variations in the initial slope, $\eta_0$, do not affect the final slope value,
although the high mass end evolves differently or reaches equilibrium at different timescales.
A distribution with less dust initially ($\eta_0 < 1.87$) also takes more time to reach equilibrium. 
A shallow distribution with an initial slope of $\eta_0 = 1.5$ takes as much as $\sim1~{\rm Gyr}$ 
to reach equilibrium, although such initial distribution slopes are unlikely. As shown by \cite{lohne08}, 
the evolution of the particle mass distribution is scalable by the total mass (or number densities of 
particles), which is what we see in the bottom two panels of Figure \ref{fig:etaM}. All systems with 
different initial masses reach the same equilibrium mass distribution, but on different timescales. 
More massive systems evolve on shorter timescales, thus reaching their equilibrium more quickly, 
while less massive systems evolve more slowly.	

\subsection{The dependence of the quasi steady-state distribution function on the collision parameters}

The parameters that describe the outcomes of collisions, in principle, should be roughly the same 
for all collisional systems. They are the fragmentation constants and the parameters of the strength 
curve \citep{benz99}. To investigate their effects on the evolution of the particle mass distribution, 
we vary their nominal values and evolve the models to the same $10~{\rm Gyr}$, as we did for the 
reference model. 

We give here a detailed analysis of the effects of varying only five of the twelve parameters ($\alpha$, 
$b$, $Q_{\rm sc}$, $s$, $S$), as the remaining seven ($\gamma$, $\beta_{\rm X}$, $f_{\rm Y}$, 
$f_{\rm X}^{\rm max}$, $f_{M}$, $g$, $G$) have no significant effects (see Table \ref{tab:tabvar} for 
the description of these parameters). 

\begin{figure*}[!ht]
\begin{center}
\includegraphics[angle=0,scale=1.4]{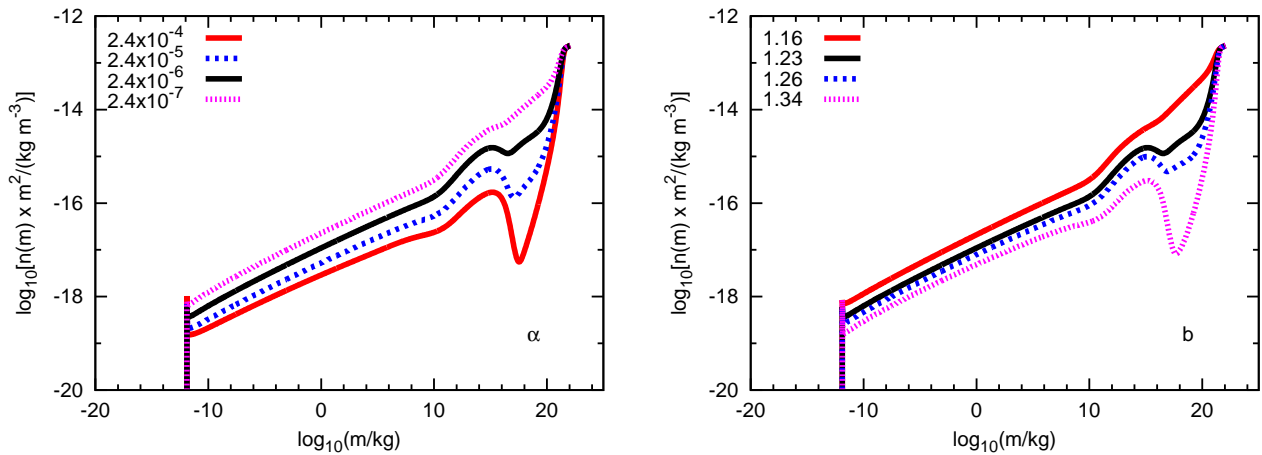}
\caption{Particle mass distribution at $10~{\rm Gyr}$, when varying the parameters $\alpha$ (left)
and $b$ (right) of the cratered mass (equation \ref{eq:mcr}).
Variations in the particle distribution only occur at large masses, the dust mass-distributions are the same
for all models.}
\label{fig:alphab}
\end{center}
\end{figure*}

In Figure \ref{fig:alphab}, we plot the resulting mass distributions when varying the cratered 
mass parameters $\alpha$ and $b$ \citep{koschny01a, koschny01b}. The parameter $\alpha$ is the 
total scaling and $b$ is the exponent of the projectile's kinetic energy in the equation of the cratered mass
\begin{equation}
M_{\rm cr} = \alpha \left(\frac{\mu V^2}{2}\right)^b\;.
\label{eq:mcr}
\end{equation} 
As can be seen in Figure \ref{fig:alphab}, the resulting mass distributions depend on the values of 
$\alpha$ and $b$ only in the gravity dominated regime. At these larger masses, our model is incomplete, 
because we do not include aggregation. When increasing $\alpha$, i.e., basically softening the materials 
or increasing the effects of erosions, the number of eroded particles in the gravity-dominated regime increases 
rapidly. A similar effect can be observed when increasing the value of $b$. However, within reasonable 
values of $\alpha$ and $b$, the variation of the equilibrium particle mass distribution slope in the dust 
mass regime is negligible.

\begin{figure*}[!t]
\begin{center}
\includegraphics[angle=0,scale=2.1]{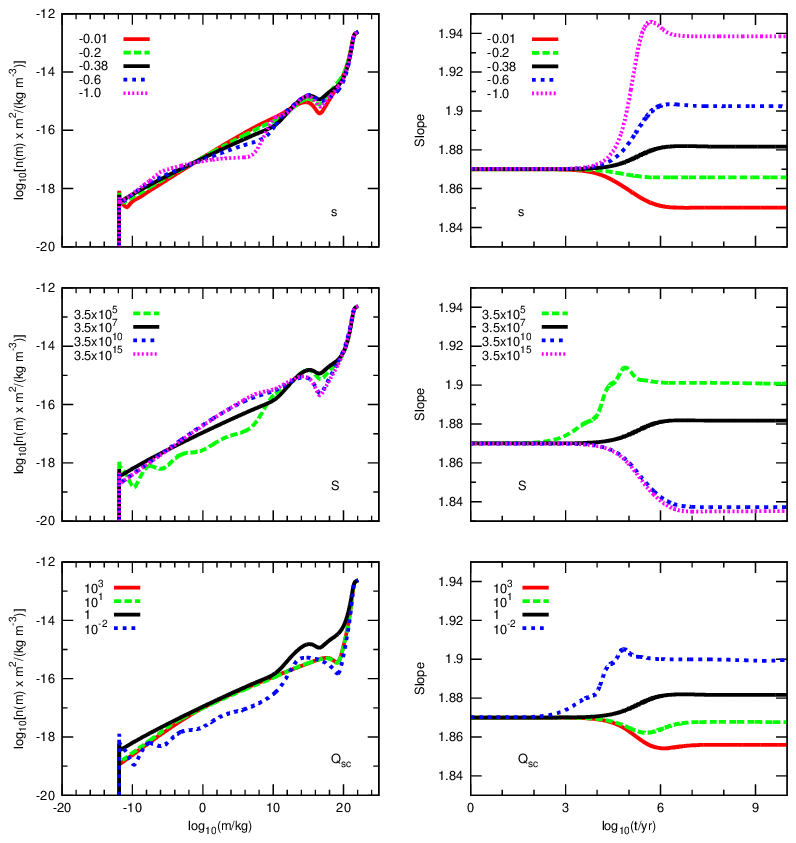}
\caption{({\it Left panels}) Particle mass distribution at $10~{\rm Gyr}$, when varying the values of the 
tensile strength curve parameters $Q_{\rm sc}$, $S$ and $s$. ({\it Right panels}) Evolution of the 
dust-mass distribution slopes when varying the values of the same parameters.}
\label{fig:QSs}
\end{center}
\end{figure*}

In Figure \ref{fig:QSs}, we plot the resulting distributions and the evolution of the dust distribution slope
when we vary the parameters $Q_{\rm sc}$, $S$, and $s$. These are all variables in the tensile strength 
curve, which is given as \citep{benz99}
\begin{equation}
Q^{\ast}(a) = 10^{-4} Q_{\rm sc}\left[S \left(\frac{a}{{\rm cm}}\right)^{s} + G \rho \left(\frac
{a}{{\rm cm}}\right)^{g}\right] \frac{{\rm J~g}}{{\rm erg}{\rm~kg}}\;.
\label{eq:BA99}
\end{equation}
The variable $Q_{\rm sc}$ is a global scaling factor, $S$ is the scaling of the strength-dominated regime,
$s$ is the power dependence on particle size of the strength-dominated regime, $G$ is the scaling of the 
gravity-dominated regime, and $g$ is the power dependence on particle size of the gravity-dominated 
regime. Variations in the gravity-dominated regime of the curve ($G$ and $g$) do not have significant 
effects on the equilibrium dust-mass distribution, so we do not consider these parameters further.
The tensile strength curve has been extensively studied for decades. However, as it is dependent 
on various material properties and the collisional velocity \citep{stewart09,leinhardt12}, its parameters do 
not have universally applicable values. Determining the tensile strength curve at large and small sizes is 
also extremely difficult experimentally. 

The slope $s$ of the strength curve in the strength-dominated regime depends on the Weibull flaw-size 
distribution. Its measured values range anywhere between -0.7 and -0.3 \citep{holsapple02}. Steeper 
values of $s$ make smaller materials harder to disrupt, which results in a steeper dust distribution slope. 
At $s=1.0$, the smallest particles are hard enough to resist catastrophic disruption even when the projectile 
mass equals the target mass. This results in a mass distribution with a slope equal to the redistribution
slope $\gamma=1.83$ at the smallest scales, while in the fitted region it is significantly steeper. At $s=0.6$, 
the smallest particles are still able to destroy each other and generate a dust distribution slope that is close to 1.91.

The scaling constants of the tensile strength curve are the dominant parameters in the evolution toward the 
quasi steady-state distribution. When reducing the complete tensile strength curve scaling $Q_{\rm sc}$, wave 
structures form more easily, as a particle becomes capable of affecting the evolution of particles much 
larger than itself (see Paper I). When upscaling the tensile strength curve, the quasi steady-state distribution slope 
starts to resemble the redistribution slope, as it is the particle redistributions that lead the evolution of the 
particle mass distribution. When varying the scaling of only the strength side of the curve $S$, similar effects 
can be seen.

We have shown that drastic offsets in the collision parameter values result in slight changes to the quasi
steady state mass distribution slope. We conclude that, for all reasonable values of the collisional parameters, 
the quasi steady-state dust-mass distribution slope is larger or equal to $1.88$.

\subsection{The dependence of the distribution function on system variables}

There are a number of parameters that can change from one collisional system to another: the material 
density $\rho$, the minimum and the maximum particle mass in the system $m_{\rm min}$ and $m_{\rm max}$, 
the radial distance $R$, height $h$, and width $\Delta R$ of the disk, and the spectral type of the central star. 
All these parameters affect three properties of the collisional model: the blow-out mass, the collisional velocity, 
and the number density of particles. Varying these parameters will change the timescale of the evolution and 
affect the quasi steady-state distribution slope. In this subsection, we analyze the effects of varying the radial
distance on the equilibrium mass distribution through the dependence of the collisional velocity on radius. Modifying either 
disk parameters $\Delta R$ and $h$ or the spectral-type of the star would have similar effects. We defer discussion 
of the variations in the timescales to a future paper.

\begin{figure*}[!t]
\begin{center}
\includegraphics[angle=0,scale=1.4]{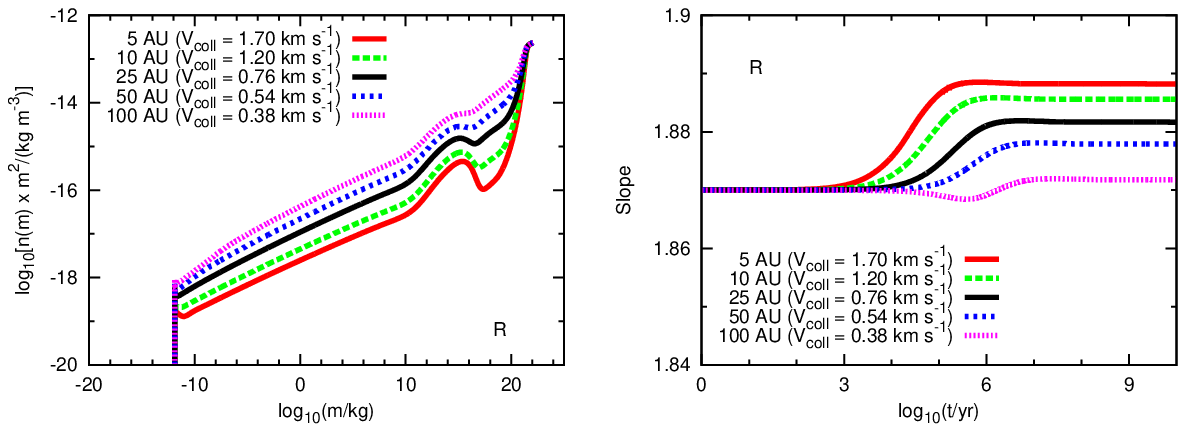}
\caption{({\it Left panel}) Particle mass distribution at $10~{\rm Gyr}$, when varying the value of 
the radial distance of the disk. ({\it Right panel}) Evolution of the dust-mass distribution slopes when varying the 
same parameter.}
\label{fig:R}
\end{center}
\end{figure*}

In the left panel of Figure \ref{fig:R}, we show the effects of varying the radial distance, $R$, on the mass 
distribution evolved to 10 Gyr. Decreasing the radial distance will increase the collisional velocity, resulting in the appearance 
of waves at the small-mass end of the distribution. It also generates a much more pronounced kink at the 
high mass end. On the other hand, when the velocity is decreased at large radii, the low mass end of the 
distribution starts resembling the redistribution function, as smaller particles are not destroyed due to the
lower energy collisions. Moreover, no kink is produced at the high mass end. In the right panel of 
\mbox{Figure \ref{fig:R}}, we show the evolution of the particle-mass distribution slope as a function of the 
collisional velocity. For high velocity collisions, the waves render the fitting of a single mass distribution
slope ill constructed but the underlying slope of the wavy mass distribution is slightly steeper than for the
smaller collisional velocity case.

\subsection{The dependence of the distribution function on numerical parameters}
\label{sec:var}

\begin{figure*}[!t]
\begin{center}
\includegraphics[angle=0,scale=1.4]{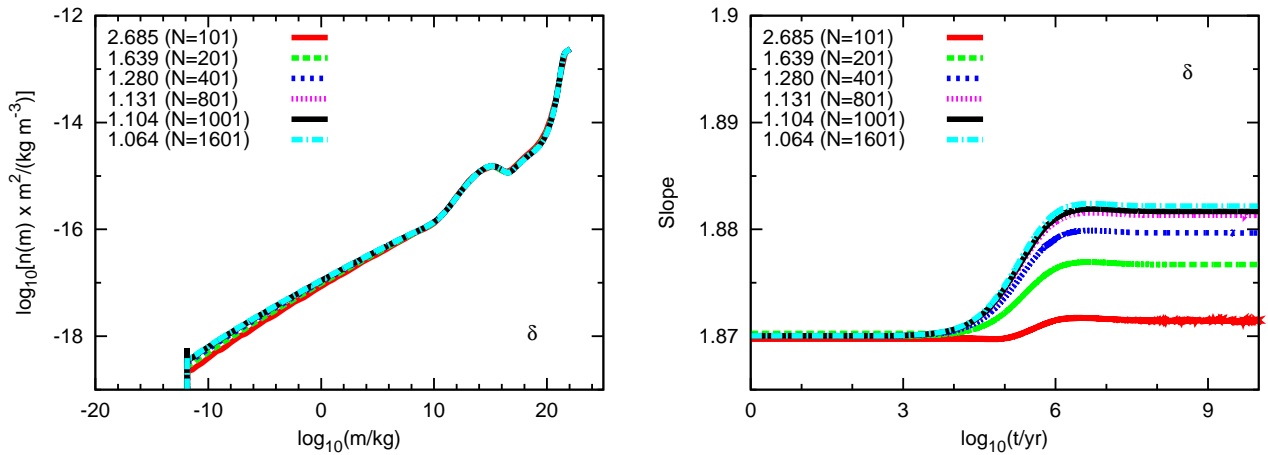}
\caption{({\it Left panel}) Particle mass distribution at $10~{\rm Gyr}$, when
varying the resolution of the numerical model. ({\it Right panel}) Evolution of the dust-mass distribution 
slopes when varying the same parameter.}
\label{fig:delta}
\end{center}
\end{figure*}

We discuss here the effects of three non-physical variables that appear in the numerical algorithm. 
They are: the mass ratio $\delta$ between neighboring grid points and the parameters of the large 
particle collisional cross-section smoothing formula, $\Theta$ and $p$ (see Paper I). In Figure \ref{fig:delta}, 
we plot the distribution of the model with varying values of $\delta$ at 10 Gyr (left panel) and the evolution 
in the slope of the dust distribution (right panel). The evolution of the dust distribution is affected by the 
number of grid points we use, converging at $\delta = 1.13$; this corresponds to 801 grid points between 
our $m_{\rm min}$ and $m_{\rm max}$ mass range. Using a lower number of grid points leads to errors 
in the numerical integration for the redistribution, leading to an offset larger than 7\% for the smallest 
particles in the system, when only using half as many grid points. We find that the dust distribution slope 
is practically independent of the smoothing variables $\Theta$ and $p$.

\subsection{The time to reach quasi steady-state}

In our model calculations, the dust distributions in the vast majority of cases reach quasi steady-state by 
\mbox{10-20 Myr} and only in a few cases do they take somewhat longer. The characteristic time is 
less than 100 Myr for all realistic cases in agreement with previous studies \citep[e.g.,][]{wyatt07,lohne08}.
This shows that, apart from second generation debris disks, 
the majority of debris disks around stars of ages over 100 Myr are most likely to be in collisional quasi
steady-state, at least for the smallest particles ($< 1~{\rm cm}$). However, young and extended systems, 
such as $\beta$ Pic \citep{smith84,vandenbussche10}, might not be near quasi steady-state at the outer 
parts of their disks, where interaction timescales are longer.

\subsection{The robustness of the solution}

\begin{figure*}[!t]
\begin{center}
\includegraphics[angle=0,scale=0.7]{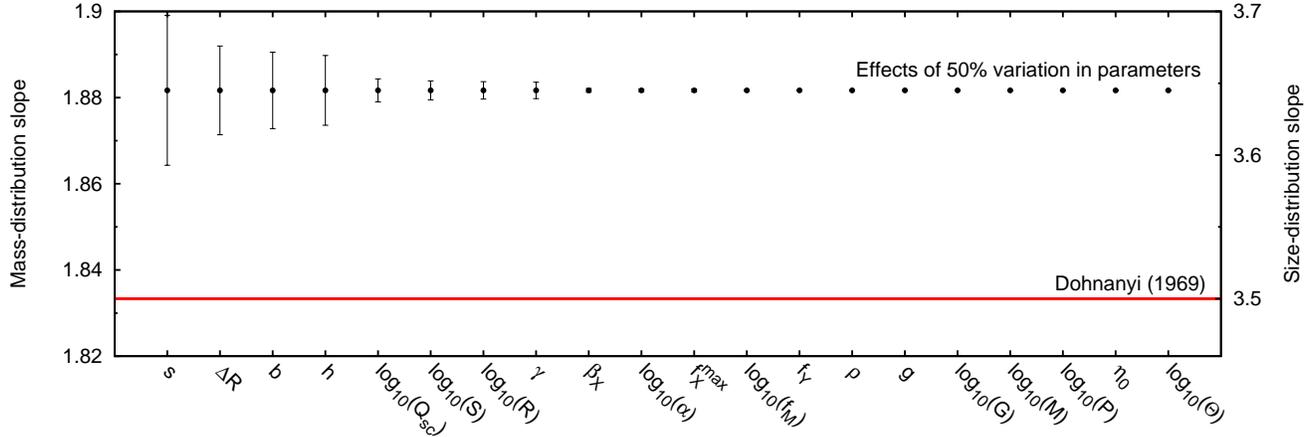}
\caption{Effect on the equilibrium particle mass and size distributions of varying each collisional and 
system variable by 50\%.}
\label{fig:errs}
\end{center}
\end{figure*}

One of the most surprising results of the wide range of numerical models we computed is the robustness 
of the quasi steady-state distribution. Varying the values of the model parameters does not result in significant 
changes in the slope of the distributions. In Figure \ref{fig:errs}, we show the effect on the equilibrium slope of 
the dust-mass distribution function of varying each model parameter. We order the parameters 
on the horizontal axis as a function of decreasing magnitude of their effects. Parameters that had to
be varied by many orders of magnitude to have effects on the equilibrium slope are analyzed in log
space. The plot shows that the dominant parameter, by far, is the slope of the strength curve in the strength-dominated 
regime. This is followed by variables that affect the collisional velocity ($\Delta R$ and $h$) and the power $b$ of the
erosive cratered mass formula (equation [\ref{eq:mcr}]). The plot also shows that neither of our arbitrarily 
chosen collision prescription constants have any significant effects on the outcome of the collisional 
cascade. The model runs predict an equilibrium dust mass distribution slope of $\eta = 1.88\pm0.02$ 
($\eta_a = 3.65\pm0.05$), taking the maximum offsets originating from the 50\% variation in the model 
parameters test as our error.

\section{The power-law approximation of the mass-distributions}\label{sec:approx}

Due to the complexity of full numerical models and the varying amount of structures in the particle
mass-distributions presented in the papers detailing them, it is still common practice for debris disk
SEDs to be calculated and fitted to observed data assuming a simple power-law for the 
mass-distribution of particles. The applicability of a power-law distribution has never been probed systematically. Another issue within this
simplification is that the value for the slope of the power-law used is generally the traditional
11/6 (or 3.5 in size space) determined by \cite{dohnanyi69}, although this has already been proven
to be only valid for the constant material tensile strength case by \cite{durda97} and \cite{obrien03}.
As we have shown in the preceding sections, a steeper distribution function of $\eta = 1.88\pm0.02$ 
($\eta_a = 3.65\pm0.05$) is a more likely outcome when considering realistic conditions. Similar
results have been found by \cite{durda97,obrien03} and \cite{thebault07}, among others.

In this section, we investigate the extent to which a simple power-law is a valid approximation 
for the mass distribution in debris disks. As we are evaluating the detectability of effects of deviations
from a power-law mass distribution on the generated SEDs, we are only concerned with particles smaller
than $\sim 1~{\rm mm}$ in size. Deviations from a power-law on this scale occur when waves appear
superposed on the distribution itself, meaning the conditions that result in the formation of waves in 
the particle mass-distribution need to be studied. 

\subsection{Appearance of waves in the particle mass-distribution}

\begin{figure*}[!t]
\begin{center}
\includegraphics[angle=0,scale=1.4]{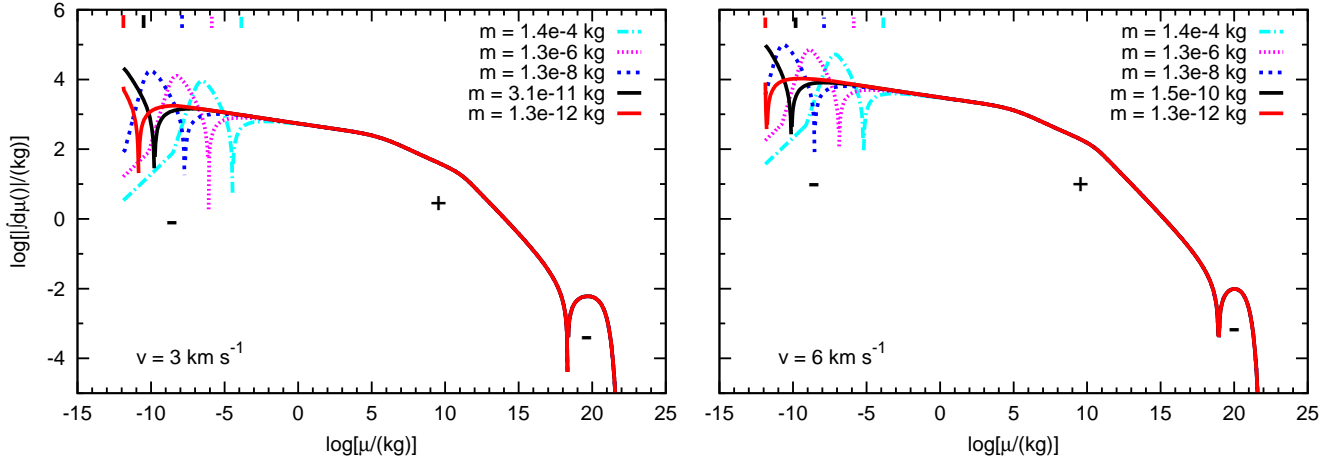}
\caption{The integrands of Equation \ref{eq:start} for selected $m$ masses in system with collisional velocities of 3 km s$^{-1}$ ({\it left panel})
and 6 km s$^{-1}$ ({\it right panel}). The solid black lines give the integrands for the mass at the first wave dip in each system ($m_{\rm wave}$), while the solid red 
lines give the integrands for the minimum mass in the system. The removal peak of the $m_{\rm wave}$ integrand is at the cut-off, explaining the excess removal
at the first dip in the waves. The tick marks show the location of the masses in the distribution for which integrands are displayed. See text for detailed description 
of the formation of the wave structure.}
\label{fig:waves}
\end{center}
\end{figure*}

Waves appear superposed on the mass-distribution due to the discontinuity at the low-mass cut-off.
This behavior has been explained before as a result of the build-up of particles at the boundary \citep{campo94,wyatt11},
however examining the properties of the collisional equation (see Paper I) yields a different picture. In Figure 
\ref{fig:waves}, we plot the $d\mu$ integrand of the collisional equation
\begin{eqnarray}
\frac{{\rm d}n(m)}{{\rm d}t} &\propto& \int_{m_{\rm min}}^{m_{\rm max}} {\rm d}\mu \Biggl\{ - m^{-\eta} \mu^{-\eta} \left(m^{\frac{1}{3}} + \mu^{\frac{1}{3}}\right)^2 \Biggr. \nonumber \\
			     && + \mu^{-\eta}M^{-\eta}\left(\mu^{\frac{1}{3}} + M^{\frac{1}{3}}\right)^2 \delta\left[X(\mu,M)-m\right] \nonumber \\
			     && + \int_{\mu}^{m_{\rm max}} {\rm d}M \mu^{-\eta}M^{-\eta}\left(\mu^{\frac{1}{3}} + M^{\frac{1}{3}}\right)^{2} \times \nonumber\\
			     &&\Biggl. A\left(\mu,M\right)H\left[Y\left(\mu,M\right)-m\right]\Biggr\}\;,
\label{eq:start}
\end{eqnarray}
which basically gives the amount of variations produced by a $\mu$ mass projectile for a certain 
$m$ mass, either by $\mu$ removing $m$ or adding it as an $X$ fragment. In regimes marked with a ``$+$'' the 
effective changes in the number densities are positive, while in regimes marked with a ``$-$'' they are negative. With 
solid black lines we show the integrands for the $m$ masses
where the first dip of the waves are situated for the corresponding velocities. It is apparent that for these masses
the peaks of the removal integrands are located exactly at the cut-off. 
The fact that they are peaks is obvious when comparing with the integrands of larger masses. If the distributions 
were continuous, this effect would smooth out, but with the cut-off located at the peak of the integrand, relatively more 
will be removed of mass $m$ at the first dip than of other masses. We also note that the additive peaks of the minimum (cut-off) mass, shown with 
red solid curves, are not
correlated with the location of the dip in the distribution (marked with a black solid tick mark), but shift as a function
of the collision velocity. At higher collisional velocity, where waves are stronger, the largest number of cut-off particles
are actually produced by the particles in the dip, which would provide a negative feedback according to the traditional
picture, canceling the waves. Our analysis shows that positive feedback is not the dominant effect in the production
of the waves.

The location of the peak of the removal term (Term I in Paper I) is the determining factor in the formation
of the waves, which can be given by solving
\begin{eqnarray}
0 & = &\partial\left\{\mu^{-\eta}\left[m^{-\eta}\left(\mu^{\frac{1}{3}}+m^{\frac{1}{3}}\right)^2 \right.\right. - \nonumber \\
&& \left.\left.\left. M^{-\eta}\left(\mu^{\frac{1}{3}}+M^{\frac{1}{3}}\right)^2 \right]\right\}\Biggl/\Biggr.\partial \mu\right |_{\mu=m_{\rm min}}\;,
\end{eqnarray}
where $m$ gives the location of the first dip and $M$ is given so that $X(\mu,M)\equiv m$. Unfortunately,
an analytic solution can not be given, as the derivative is transcendent, but is easily solvable numerically. The variable
that determines the solution will be $X$, which depends on the collisional velocity, the tensile strength law, and the 
value of $m_{\rm min}$. For a given system, where $m_{\rm min}$ and the tensile strength law do not change as
a function of radial location, the single variable determining the wavy structure is the collisional velocity. The minimum
mass will change as a function stellar types and some material/velocity dependence of the tensile strength law can
be expected between systems.

\subsection{Constraining the conditions for the appearance of waves in the particle mass-dsitribution}

\begin{figure}[!t]
\begin{center}
\includegraphics[angle=0,scale=0.7]{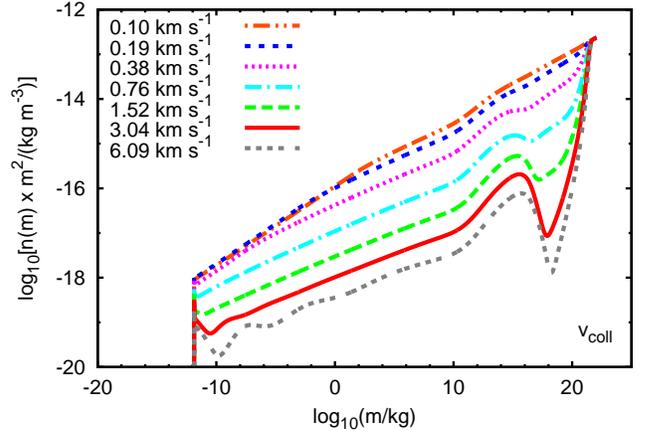}
\caption{Particle mass distribution at $10~{\rm Gyr}$, when varying the collisional velocity in the system. Waves start to appear at and above collisional velocities 
of $3~{\rm km s}^{-1}$, when constraining all other parameters at their fiducial values. {\it Note: The collisional velocity was adjusted by hand in these models and 
not determined by the system geometry as described in Paper I, in order to study the effects of only variations in the collisional velocity itself.}}
\label{fig:v}
\end{center}
\end{figure}

In Section \ref{models}, we have shown that variations in the proxies for the collisional 
velocity (such as disk radius) can in fact result in wavy size distributions, even within a narrow debris ring.
Extended debris disks will be even more likely to have higher collisional velocities, as the particles with higher
$\beta$ values (but less than 0.5) get placed on high eccentricity orbits. These small particles from the 
inner rings will collide with the particles in the external rings with increased velocities due to the non-zero
collisional angles \citep{thebault07}. To study the effects of variations only in the collisional velocity, we set the collisional velocity
directly within our code to specific values, not changing any other parameters. We present the results
from these runs in Figure \ref{fig:v}. The figure shows that waves start to appear at collisional velocity 
values of $3~{\rm km~s}^{-1}$ and above.

The value of the orbital velocity for a particle on an elliptical orbit originating from 
$R_{\rm in}$ with a certain $\beta$ value at $R_{\rm out}$ radius (where it will collide with a particle on a 
circular orbit) is
\begin{equation}
v_{\rm e} = \sqrt{{\rm G}M_{\ast}\left(1-\beta\right)\left[\frac{2}{R_{\rm out}}-\frac{1-2\beta}{\left(1-\beta\right)R_{\rm in}}\right]}\;.
\end{equation}
The angle between the orbital velocities can be expressed as
\begin{equation}
\zeta = \cos^{-1}\left[\frac{\left(2ae\right)^2-R_{\rm out}^2-\left(2a-R_{\rm out}\right)^2}{2R_{\rm out}\left(R_{\rm out} - 2a\right)}\right]\;
\end{equation}
where
\begin{eqnarray}
a&=&R_{\rm in}\frac{1-\beta}{1-2\beta}\\
e&=&\frac{\beta}{1-\beta}\;.
\end{eqnarray}
The collisional velocity is then
\begin{equation}
v_{\rm coll} = \sqrt{\left(v_{\rm o}-v_{\rm e}\right)^2 + \left(v_{\rm e}\sin\zeta\right)^2}\;,
\end{equation}
where $v_{\rm o}$ is the orbital velocity for the particle on a circular orbit at $R_{\rm out}$.

\begin{figure*}[!t]
\begin{center}
\includegraphics[angle=0,scale=1.4]{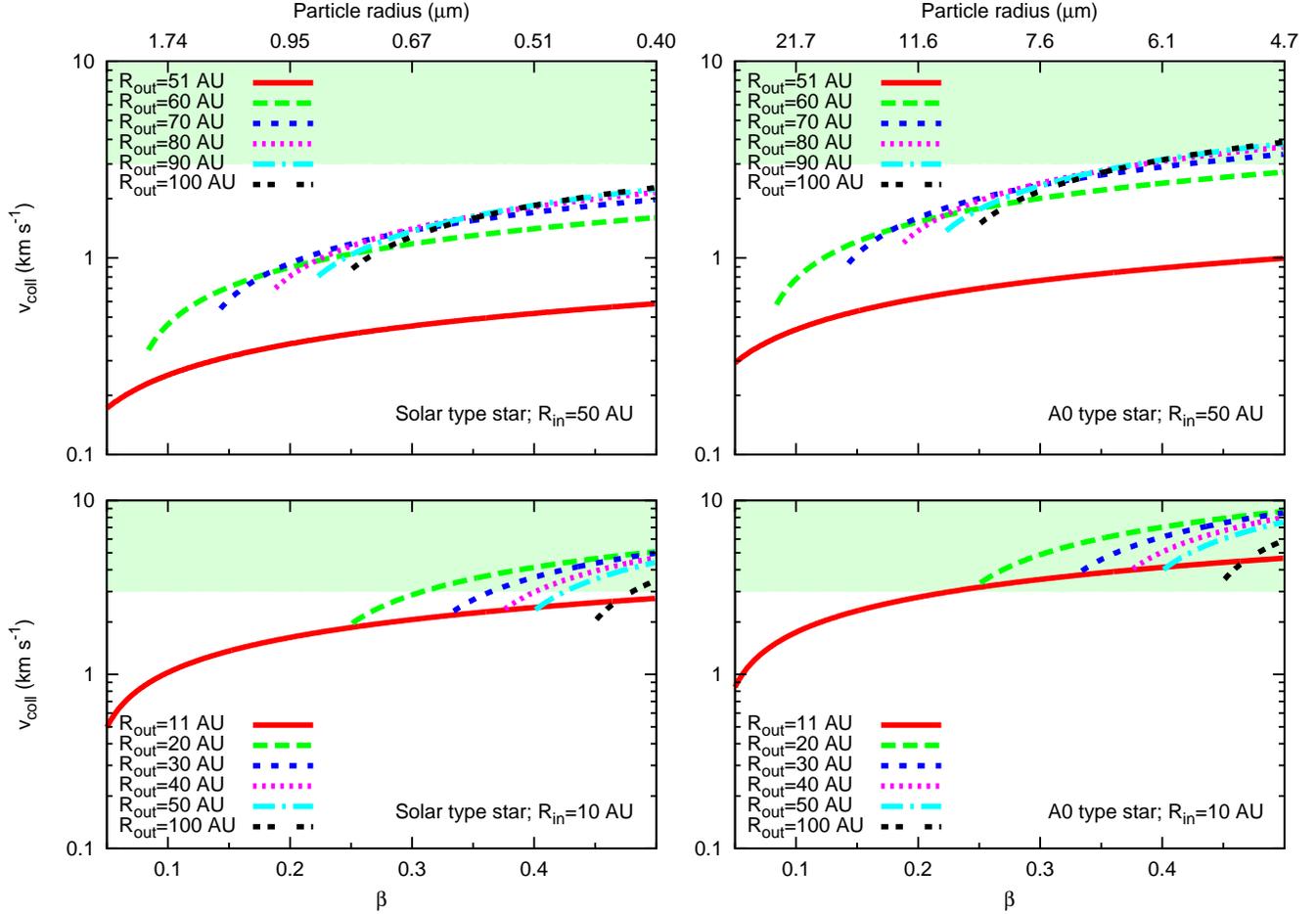}
\caption{The collisional velocity between particles originating from a radius of $R_{\rm in}$ and colliding with particles at radius $R_{\rm out}$, with eccentricities
calculated from their $\beta$ values. We show cases for a solar-type star and an A0 spectral-type star, with $R_{\rm in}$ values of 10 and 50 AU.}
\label{fig:vcol}
\end{center}
\end{figure*}

We plot the values of this collisional velocity for particles originating from $R_{\rm in}=10~{\rm AU}$
and $50~{\rm AU}$ in systems around a solar- and early-type star in Figure \ref{fig:vcol}. We shade the
collisional velocity region above our $3~{\rm km~s}^{-1}$ limit, where waves start to appear. As we can 
see, nearby rings at $\Delta R=1~{\rm AU}$ have low collisional
velocities, reassuring that our prescription for a constant collisional velocity value in narrow debris rings is
reasonable. We can also see that once we depart from the narrow ring assumption 
($\Delta R > 10~{\rm AU}$), the collisional velocities do increase.

Around a solar-type star, particles originating
from an inner radius of $10~{\rm AU}$ are able to achieve collisional velocities of $\sim 3~{\rm km~s}^{-1}$;
however, the particle size range that is able to achieve this limit is very narrow, between 0.4 and $0.6~\micron$.
Taking into account the dilution of these small particles within a limited narrow size range, we conclude that 
extended debris disks with inner boundaries outside of 10 - $15~{\rm AU}$ around solar type stars should be
well approximated by a simple power-law mass (size) distribution function.
In Paper I, we show a comparison run to \cite{lohne08} and \cite{wyatt11}, assuming an extended disk
between radii of 7.5 and $15~{\rm AU}$ around a solar-type star. The runs by \cite{wyatt11} and our code
show negligible/small amplitude waves, while \cite{lohne08} show moderate amplitude waves, likely due to
their full three dimensional modeling of the system, taking into account smaller particles originating from
regions inward of 10 AU. These results confirm our analytic analysis on the conditions required to initiate
waves in the dust particle mass-distribution.

We show the same analysis for early-type stars in the right panels of Figure \ref{fig:vcol}. The collisional
velocities of particles originating from $10~{\rm AU}$ with particles on external circular orbits around an 
A0 spectral-type star will be significantly larger than our limit for a larger range of particle sizes (5 - $10~\micron$). Particles
originating from orbits inside of $10~{\rm AU}$ will definitely initiate waves in extended disks or external
rings. However, particles originating from $50~{\rm AU}$ will barely reach collisional velocities at or 
above our limit and for a limited range of particle sizes. We conclude that extended debris disks with 
inner boundaries outside of 50 - $60~{\rm AU}$ around early-type stars should be well approximated by 
a simple power-law mass (size) distribution function

Our 1D ``particle-in-a-box'' numerical code adjusts the collisional velocities through the ring thickness and height ($\Delta R$ and $h$), and the resulting orbital inclinations for the particles. Our results are only moderately affected over the relatively narrow range of velocities obtained by varying them (see Figure \ref{fig:errs}). We modified the code (see above) to simulate the higher-velocity collisions with particles on elliptical orbits to determine a threshold for generation of waves in the particle distribution. However, our model cannot explore the nuances resulting from varying collisional velocities in extended systems (such as varying collisional rates and collisional outcomes), which could also play a role in the possible formation of substructures superposed on the generally steeper particle size-distributions \citep[see e.g.,][]{krivov05,thebault07,lohne08,muller10}.

As stated at the end of Section 3.1, our generalization assumed a specific system (with same material
properties and minimum particle mass). Although we explored varying the minimum particle mass when
introducing different spectral type central stars, both the tensile strength and the minimum mass may vary
with material properties as well. In addition to minerals such as basalt, it is likely that grains in the outer zones 
of debris disks contain significant amounts of ice. The mechanical properties of icy grains are explored
by \cite{benz99} and \cite{leinhardt09}, yielding roughly a factor of 3 in difference between the tensile strength
of basalt and ice. Given the relative insensitivity of our results to the scaling of the grain strength curve ($Q_{\rm sc}$ - see Figure \ref{fig:errs}), 
these properties are similar enough that our basic conclusions about the grain size distribution and region of applicability 
of the power law approximation to it are essentially unchanged even outside of the iceline.

\section{Comparison to previous analytic approaches}\label{sec:companal}

\begin{deluxetable*}{llllll} 
\tablecolumns{6}
\tabletypesize{\scriptsize}
\tablewidth{500pt}
\tablecaption{Comparing the properties of analytic solutions with our numerical model\label{tab:companal}}
\tablehead{
\colhead{Model} 	& \colhead{Collision rate}	& \colhead{$X/M$}		 	& \colhead{$Q_{D}^{\ast}$} 	& \colhead{Redistribution}	 	& \colhead{$\frac{d \ln n(m)}{d \ln m}$}} 
\startdata
\cite{dohnanyi69}	& $m^{2/3}$			& Constant				& Constant				& power-law					& 11/6 \\
\cite{tanaka96}		& $m^{\nu}$			& Constant				& Constant				& power-law					& $\left(\nu + 3\right)/2$ \\
\cite{obrien03}		& $m^{2/3}$			& Constant				& power-law				& power-law					& $\left(11+s\right)/\left(6+s\right)$ \\
\cite{belyaev11}	& $m^{2/3}$			& Variable					& power-law				& power-law					& slowly varying function of $m$ \\
\hline
\multirow{2}{*}{This work}	 & \multirow{2}{*}{variable} & \multirow{2}{*}{Variable} & \multirow{2}{*}{power-law} & \multirow{2}{*}{power-law} & Function of model parameters,\\
				&					&						&						&							& largely robust around -1.88
\enddata
\end{deluxetable*}

The slope of the size distribution slope in collisional cascades has been investigated with analytic approaches 
with increasing complexity since the pioneering work of \cite{dohnanyi69}. His assumptions of self-similar 
fragmentation, constant material strength, and collision rates proportional to the geometric cross section
lead to the well known and now widely used steady state size-distribution slope of $\eta_m = 11/6$ ($\eta_a = 3.5$).
The validity of his assumptions has been investigated in a number of analytic studies.
\cite{tanaka96} found that varying the power of the collisional cross section ($m^{\nu}$, where $\nu=2/3$ for a simple
geometric cross section) will set the steady state distribution slope as $\eta_m = (\nu+3)/2$. Such non-geometric
cross sections (collision rates) can be the results of mass weighted collision probabilities, varying amounts of grain porosity,
mass dependent collisional velocities, etc. \cite{obrien03} investigated whether a non-constant tensile strength 
value, defined by a power-law with slope of $s$, will affect the steady state particle size distribution. They found
a linear relationship between the two power-law slopes, yielding a steady state solution producing more small
particles than the traditional \cite{dohnanyi69} solution, when considering realistic tensile strength slopes. 
An analytic solution similar to the one by \cite{obrien03} was presented in \cite{wyatt11}, who also analyzed
their equations numerically, revealing substructures superimposed on the power-law size distributions.
\cite{belyaev11} expanded on the \cite{obrien03} model by allowing a variable largest fragment mass to target 
mass ratio ($X/M$ in our model), following the \cite{fujiwara77} experimental results. \cite{belyaev11} found a particle 
distribution that could not be
described by a single power-law, rather by a slowly varying, mass dependent power-law function. We summarize 
the characteristic properties of these analytic models in Table \ref{tab:companal}. A common issue with all of these 
analytic solutions is that they assume a steady state system, meaning that the mass flux through mass $m$ is zero 
[$\partial F(m) / \partial m = 0$]. In reality, without mass input at the high mass end the systems lose mass and decay, 
meaning the mass flux is never zero in a system, but rather is a time and mass dependent variable. The solution of the 
cascade is actually an eigenvalue problem, which is difficult to solve analytically.

\begin{figure}[!t]
\begin{center}
\includegraphics[angle=0,scale=0.66]{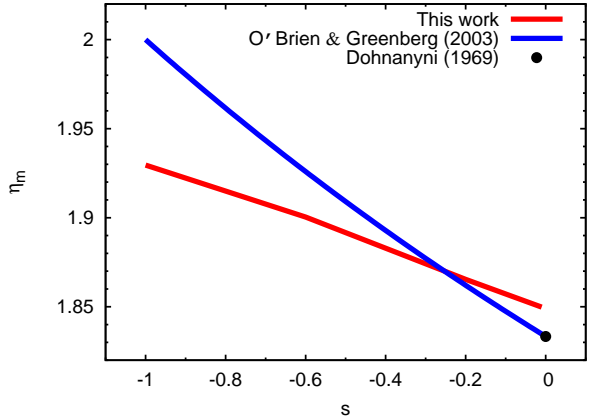}
\caption{Comparing the particle mass-distribution slope vs.\ tensile-strength curve slope of \cite{dohnanyi69}, \cite{obrien03}, and our current work.}
\label{fig:seta}
\end{center}
\end{figure}

Our numerical code allows investigating the quasi steady state [$\partial F(m) / \partial m \ne 0$]
solution of collisional cascades, without the aforementioned simplifications. In Figure \ref{fig:seta}, we compare
the \cite{dohnanyi69} and \cite{obrien03} steady state mass distribution slope results with the ones given by our 
code at varying values of tensile strength slope. At $s=0$ (constant $Q^{\ast}_D$) the \cite{obrien03}
solution is equal to the traditional \cite{dohnanyi69} value, but it gives steeper mass-distribution slopes for lower values of $s$.
Similar behavior is seen with our code; however, at $s=0$ our prediction for the quasi steady state slope
is $\eta_m=1.85$ ($\eta_a = 3.55$), which is still steeper than the traditional value. At much lower values of $s$ ($<-0.25$) 
we predict a mass distribution slope that is less steep than the ones given by the \cite{obrien03} formula, while our results
intersect at $s=-0.25$, yielding a quasi steady state slope of $\eta_m = 1.87$ ($\eta_a = 3.61$). Since the point of
this intersection is close to the theoretical and experimental value of the tensile strength curve slope, the predictions
from a pure steady state model and a quasi steady state model will agree within errors. This renders the testing of
different models difficult. However, our numerical calculations provide a general new estimate for the particle size
distribution slopes with significantly more complete physics compared to previous analytic approaches.

\section{Synthetic Spectra}
\label{sec:synthetic}

In the following sections, we compare the emission that results from the predicted particle-mass distributions 
to observations. As a first step, we generate an array of synthetic spectra using realistic astronomical silicate 
emission properties. We then analyze how the spectra are influenced by the particle mass distribution function.

The flux emitted by a distribution of particle masses at a certain frequency is equal to
\begin{equation}
F_{\nu} = \frac{{\mathfrak v}\pi}{D^2}\int_{\rm n}^{\rm x} {\rm d}a~ n(a) a^2  Q_{\rm abs}\left(a,\nu\right) B_{\nu}\left(T\right)\;,
\end{equation}
where $Q_{\rm abs}$ is the absorption efficiency coefficient, $B_{\nu}\left(T\right)$ is the 
blackbody function, and ${\mathfrak v}$ is the total volume of the emitting region. Since in infrared
astronomy it is customary to express the flux density as a function of wavelength, we rewrite this
also as
\begin{equation}
F_{\nu} = \frac{{\mathfrak v}\pi\lambda^2}{D^2c}\int_{\rm n}^{\rm x} {\rm d}a~ n(a) a^2  Q_{\rm abs}\left(a,\lambda\right) B_{\lambda}\left(T\right)\;.
\end{equation}

\begin{figure}[!t]
\begin{center}
\includegraphics[angle=0,scale=0.7]{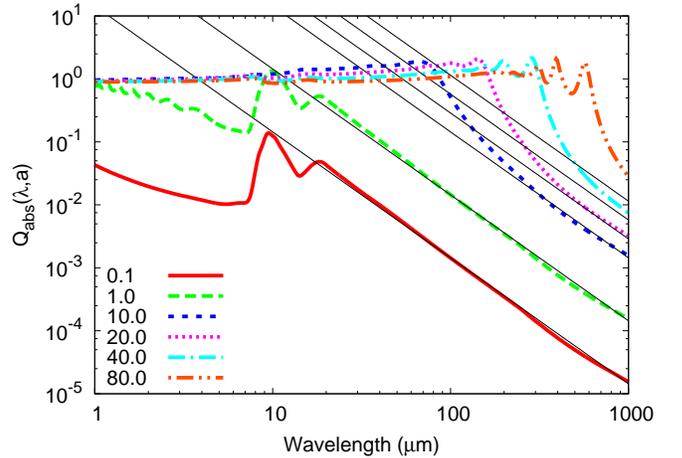}
\caption{Absorption efficiency of astronomical silicates as a function of
wavelength for a range of particle sizes between 0.1 and 80 $\micron$. 
Solid lines are the $\sim \lambda^{-\beta}$ approximations to the long-wavelength
regimes of the curves that we employ in this work.}
\label{fig:sil}
\end{center}
\end{figure}

The exact function of the absorption efficiencies of particles in the interstellar medium or in circumstellar disks is 
largely unknown. The most commonly used particle types for SED calculations are artificial astronomical silicate 
(the properties of which are adjusted to reproduce the typical 10 $\micron$ silicate feature and measured
laboratory dielectric functions) and graphite \citep{draine84}. In \mbox{Figure \ref{fig:sil}}, we plot the absorption 
efficiency as a function of wavelength for a few astronomical silicate particle sizes. Particles larger than $10~\micron$ 
have nearly constant absorption efficiency curves at shorter wavelengths ($\lambda < 2\pi a$, where $a$ is the particle 
radius) with $Q_{\rm abs}=1$, which is followed by a power-law cut off. The slope of this power-law becomes constant 
for wavelengths larger than $\sim 8\pi a$, and is commonly denoted by the variable $\beta$. Astronomical silicates of 
all sizes have a typical value of $\beta=2$ \citep{draine84}.

\begin{figure}[!ht]
\begin{center}
\includegraphics[angle=0,scale=0.7]{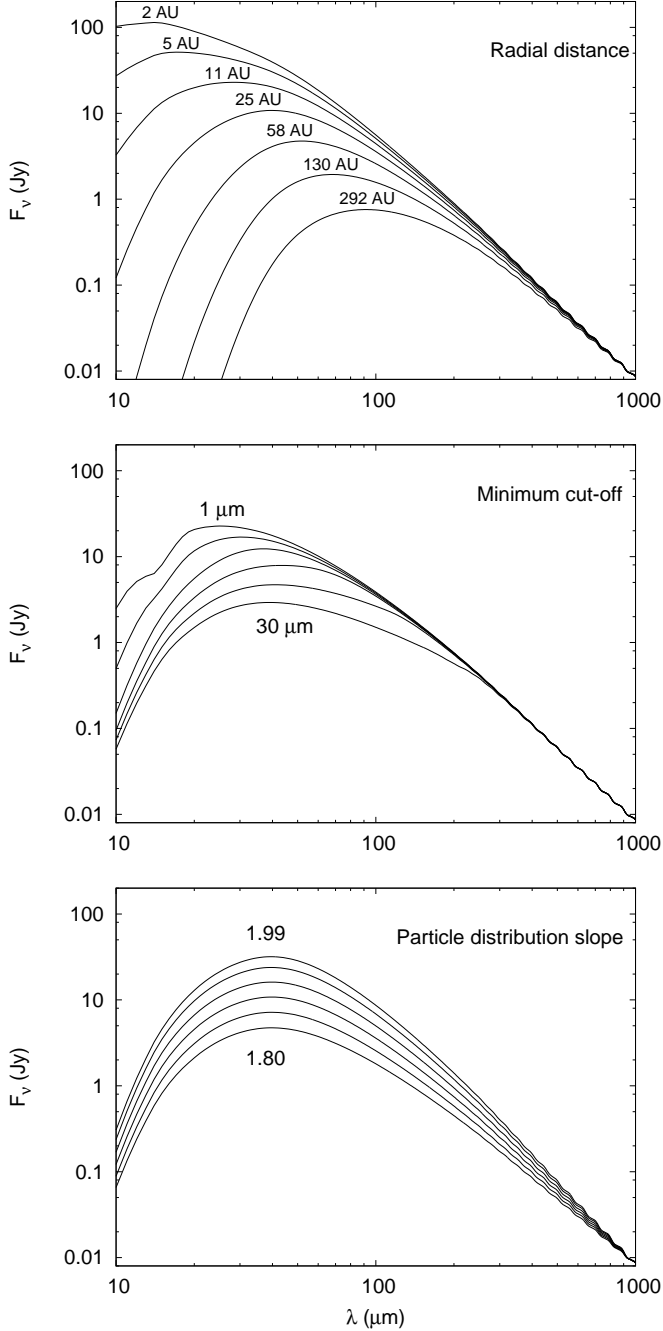}
\caption{\footnotesize{Synthetic SEDs for an array of model power-law particle mass distributions, with varied parameters. 
The fluxes are scaled to match at $1000~\micron$. In the top panel, we show the synthetic SEDs generated 
for a variety of radial distances; in the middle panel, we show synthetic SEDs generated for a variety of 
minimum cut-offs (1, 2, 4, 8, 15 and $30~\micron$); and in the bottom panel, we show synthetic SEDs 
generated for a variety of particle-mass distribution slopes (1.80, 1.84, 1.88, 1.92, 1.96 and 1.99). {\it Note:} 
In the middle-panel we vary the minimum cut-off of the particle mass distributions, even though it is a parameter
inherently set by equations in our collisional model. Since in reality, the placement of the cut-off is set by the 
optical properties and structural build of the micron size particles, it is generally treated as a variable in SED 
models. In the plot we show that such variations in the placement of the minimum cut-off do not affect the 
long-wavelength part of the SEDs.}}
\label{fig:synth}
\end{center}
\end{figure}

In Figure \ref{fig:synth}, we show synthetic SEDs, all scaled to the same flux level at $1000~\micron$. The top panel 
shows spectra calculated around an A0 spectral-type star, with debris rings placed at various distances between 2 
and $292~{\rm AU}$. The minimum particle size cut-off was set at $\sim 5~\micron$, in accordance with our model 
(see Paper I). All disks with radial distances below $\sim 130~{\rm AU}$ have a common slope for wavelengths larger 
than $250~\micron$, and the furthest disk at $292~{\rm AU}$ joins this common slope around $\sim 350~\micron$. 

The blow-out size in a system depends on grain structure (porosity) and the exact value of the optical constants
for small grains (which is largely unknown and is a function of grain material). For this reason, we also calculated 
synthetic SEDs for a debris ring at $25~{\rm AU}$ around an A0 spectral-type star, with the minimum particle size of the
distribution artificially cut off at sizes between 5 and $30~\micron$. (Note that we normally calculate the blow-out mass
self-consistently as described in Paper I.) We plot these SEDs in the middle panel of Figure \ref{fig:synth}. The offsets 
between the SEDs become apparent for wavelengths shorter than $200~\micron$, while for longer wavelengths, the 
emission profiles agree and have a common pseudo Rayleigh-Jeans slope.

Finally, we explore the dependence of the SED on the slope of the quasi steady-state particle-mass distribution. The bottom 
panel of Figure \ref{fig:synth} shows synthetic SEDs generated for a debris ring at 25 AU around an A0 spectral type 
star, with a minimum particle cut-off size at $5~\micron$, but with particle mass distribution slopes between 1.81 and 
1.99. These plots show that the slope of the Rayleigh-Jeans part of the emission is greatly influenced by the particle 
size distribution slope. In fact, it depends almost solely on this slope, with the temperature of the grains having mild effects 
at large orbital distances. We also performed our tests with dirty-ice optical constants \citep{preibisch93} and found very 
similar results, showing that our results are also independent on particle types assumed.

\section{Relation between the particle mass distribution and the SED}
\label{sec:relation}

The absorption efficiency curves can be simplified and described as
\begin{eqnarray}
Q_{\rm abs}(\lambda,a) \propto
\begin{cases}
1 & \lambda < 4 \pi a \nonumber \\
\left(\frac{x a}{\lambda}\right)^{\beta} & \lambda > 4 \pi a
\end{cases}
\end{eqnarray}
where $x$ is a scaling constant for the power-law part of the function. Fitting the silicate absorption efficiency 
functions, we find 
\begin{equation}
x = 12 \left(\frac{a}{\micron}\right)^{-0.5}\;.
\end{equation}
Using this simplified absorption efficiency model and assuming that all particles contribute to the Rayleigh-Jeans 
tail of the SED with their own Rayleigh-Jeans emission, we estimate the emitted flux density at long wavelengths as
%
%
%
\begin{eqnarray}
F_{\nu} &=&  \frac{2{\mathfrak v}\pi {\rm k}_{\rm b} T C_{\rm disk}}{D^2 \lambda^2}
\left[10^{-3}\left(\frac{12}{\lambda}\right)^{\beta} \int_0^{\frac{\lambda}{4\pi}}{\rm d}a~a^{2+\frac{\beta}{2}-\eta_a} + \right. \nonumber \\
&& \left. \int_{\frac{\lambda}{4\pi}}^{\infty} {\rm d}a~a^{2-\eta_a}\right]\;.
\end{eqnarray}
%
%
%
Here we assumed a $\beta$ parameter that is independent of the particle size. The variable $C_{\rm disk}$ is the 
number density scaling (see Paper I), ${\rm k}_{\rm b}$ is the Boltzmann constant, $T$ is the temperature of the 
dust grains (which we also assume to be particle size independent), and $D$ is the distance of the system from 
the observer. The quantity $\eta_a$ is the quasi steady-state particle size distribution slope, and can be calculated from 
the mass distribution slope as $\eta_a = 3 \eta - 2$. Integrating these functions, we get 
\begin{equation}
F_{\nu} (\lambda) = {\rm C}_1 ~ \lambda^{1-\frac{\beta}{2}-\eta_a} + {\rm C}_2 ~ \lambda^{1-\eta_a}\;,
\end{equation}
where
\begin{eqnarray}
{\rm C}_1 &=&  \frac{2{\mathfrak v}\pi {\rm k}_{\rm b} T C_{\rm disk}}{D^2} \times \frac{10^{-3}12^{\beta}2^{2\eta_a-5-\beta}\pi^{\eta_a-3-\beta/2}}{6+\beta-2\eta_a} \\
{\rm C}_2 &=&  \frac{2{\mathfrak v}\pi {\rm k}_{\rm b} T C_{\rm disk}}{D^2} \times \frac{\left(4\pi\right)^{\eta_a-3}}{\eta_a-3}\;.
\end{eqnarray}
Assuming $\beta=2$, which is appropriate for astronomical silicates, we find that the slope of the SED is equal 
to $-\eta_a$ for the short wavelength part of the Rayleigh-Jeans tail of the SED and $1-\eta_a$ for the long 
wavelength regime, where $Q_{\rm abs}(\lambda,a)=1$ for the particles contributing the most to the emission. 
Similar results have been found by \cite{wyatt02}. This behavior explains why the slope of the flux density in the long 
wavelength regime is not dependent on the optical properties of the grains, as $Q_{\rm abs}(\lambda,a)=1$ for all 
grain types at these wavelengths that effectively contribute to it.

Our models yield a quasi steady-state distribution slope of $\eta_a\approx3.65$, meaning that the Rayleigh-Jeans tail 
end of the SEDs should be proportional to 
\begin{equation}
F_{\nu} \propto \lambda^{-2.65}\;,
\end{equation}
as long as the particles are in collisional quasi steady-state.

\section{Comparison to observations}
\label{sec:constraints}

\begin{deluxetable*}{lrrrrrl}
\tablecolumns{7}
\tabletypesize{\scriptsize}
\tablewidth{500pt}
\tablecaption{Observational data of debris disks\label{tab:data}}
\tablehead{
\colhead{Star} & \colhead{$\lambda$} ($\mu$m) 	& \colhead{Flux} (mJy) 	& \colhead{Error} (mJy) 	& \colhead{Excess} (mJy) 	& \colhead{Reference} 	& Notes } 
\startdata
$\beta$ Pic 	& 250 						& 1,900.0	 			& 285.0 				& 1,897.5				& \cite{vandenbussche10}	&	\\
			& 350 						& 720.0	 			& 108.0 				& 718.7				& \cite{vandenbussche10}	&	\\
			& 500 						& 380.0	 			& 57.0 				& 379.4				& \cite{vandenbussche10}	&	\\
			& 800						& 115.0				& 30.0				& 114.8 				& \cite{zuckerman93}		&	\\
			& 850						& 85.2				& 13.0				& 85.0				& \cite{holland98}			&	\\
			& 850						& 104.3				& 16.0				& 103.8				& \cite{holland98}			&	\\
			& 1200						& 24.3				& 4.0				& 24.2				& \cite{liseau03}			&	\\
			& 1200						& 35.9				& 5.0				& 35.8				& \cite{liseau03}			& within 40" \\
			& 1300						& 24.9				& 4.0				& 24.8				& \cite{chini91}				& 	\\
\hline
$\epsilon$ Eri	& 350						& 366.0				& 109.8				& 359.45				& \cite{backman09}			& CSO/SHARCII \\			
			& 450						& 250.0				& 75.0				& 246.06				& \cite{greaves05}			& JCMT/SCUBA \\
			& 850						& 37.0				& 5.55				& 35.92				& \cite{greaves05}			& JCMT/SCUBA \\
			& 1300						& 24.2				& 4.0				& 23.74				& \cite{chini91}				& JCMT/SCUBA \\
\hline
Fomalhaut	& 350						& 1,180.0				& 354.0				& 1,168.3				& \cite{marsh05}			& \\
			& 450						& 595.0				& 200.0				& 587.8				& \cite{holland03}			& \\
			& 850						& 97.0				& 14.55				& 95.1				& \cite{holland03}			& \\
			& 1300						& 21.0				& 3.5				& 20.17				& \cite{chini91}				& within 24"\\
			& 7000						& 0.4				& 0.065				& 0.37				& \cite{ricci12}				& \\
\hline
HD 8907		& 450						& 22.0				& 11.0				& 21.87				& \cite{najita05}			& \\
			& 850						& 4.8				& 1.2				& 4.76				& \cite{najita05}			& \\
			& 1200						& 3.2				& 0.9				& 3.18				& \cite{roccatagliata09}		& \\
\hline
HD 104860	& 350						& 50.1				& 15.0				& 50.0				& \cite{roccatagliata09}		& \\
			& 450						& 47.0				& 14.0				& 46.9				& \cite{najita05}			& \\
			& 850						& 6.8				& 1.2				& 6.78				& \cite{najita05}			& \\
			& 1200						& 4.4				& 1.1				& 4.39				& \cite{roccatagliata09}		& \\
			& 3000						& 1.35				& 0.67				& 1.35				& \cite{carpenter05}			& \\
\hline
HD 107146	& 350						& 319.0				& 90.0				& 318.8				& \cite{roccatagliata09}		& \\
			& 450						& 130.0				& 39.0				& 129.9				& \cite{najita05}			& \\
			& 850						& 20.0				& 3.2				& 19.96				& \cite{najita05}			& \\
			& 1300						& 10.4				& 3.0				& 10.39				& \cite{najita05}			& \\
			& 3000						& 1.42				& 0.3				& 1.41				& \cite{carpenter05}			& \\
\hline
HR 8799		& 350						& 89.0				& 26.0				& 88.8				& \cite{patience11}			& \\
			& 850						& 15.0				& 3.0				& 14.96				& \cite{williams06}			& Aperture Correction\\
			& 1200						& 4.0				& 2.7				& 3.98				& \cite{bockelee94}			& \\
\hline
Vega		& 250						& 1,680.0				& 260.0				& 1,617.6				& \cite{sibthorpe10}			& \\
			& 350						& 610.0				& 100.0				& 578.5				& \cite{sibthorpe10}			& \\
			& 500						& 210.0				& 40.0				& 194.8				& \cite{sibthorpe10}			& \\
			& 850						& 45.7				& 7.0				& 40.5				& \cite{holland98}			& \\
\hline
HD 207129	& 250						& 113.0				& 18.0				& 111.78				& \cite{marshall11}			& \\
			& 350						& 44.3				& 9.0				& 43.68				& \cite{marshall11}			& \\
			& 500						& 25.9				& 8.0				& 25.60				& \cite{marshall11}			& \\
			& 870						& 5.1				& 2.7				& 5.00				& \cite{nilsson10}			& 
\enddata
\end{deluxetable*}

To compare the computed spectra of quasi steady-state collisional disks to data, we assembled the available data for 
debris disks with far-IR and submillimeter observations. As a result of our analysis in \S \ref{sec:synthetic}, where 
we determined the wavelength range that is least sensitive to parameters, we use only data at wavelengths larger 
than $250~\micron$. To fit a power-law to the Rayleigh-Jeans regime of the SEDs, we need a minimum of three 
data points above our wavelength cut-off. We found a total of only nine sources that fulfill these requirements. We 
present the far-IR/submillimeter fluxes for these sources in \mbox{Table \ref{tab:data}}. Occasionally, published 
submillimeter measurements do not account for systematic errors. In these cases, we applied a total of 30\%
error to all ground based measurements at 350 and $450~\micron$ and 15\% for all Herschel data and 
measurements above $850~\micron$. We also made sure that the data included all the flux from each source 
and applied an aperture correction estimate otherwise. All corrections are listed as notes in Table \ref{tab:data}.

\begin{figure*}[!ht]
\begin{center}
\includegraphics[angle=0,scale=0.76]{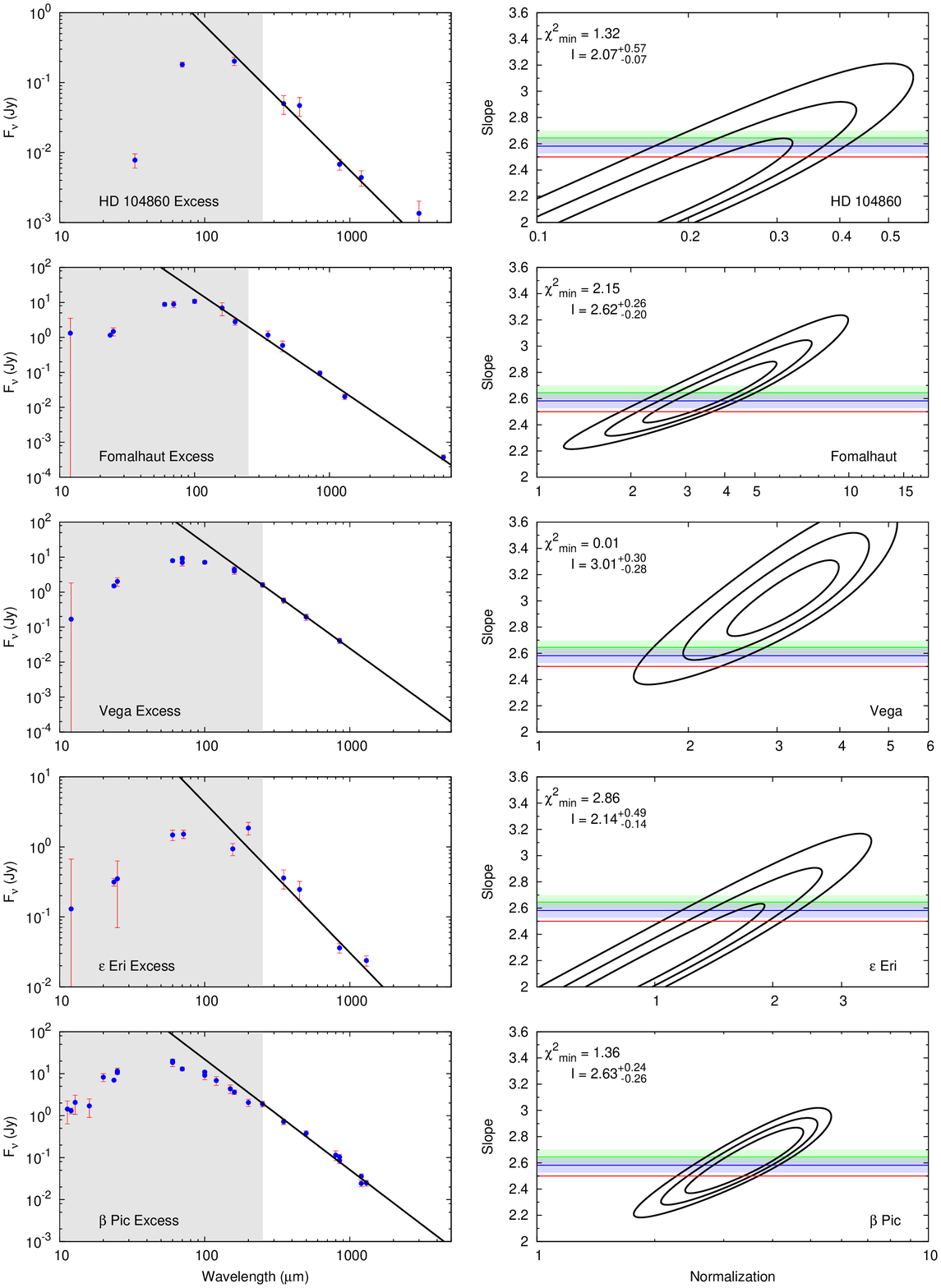}
\caption{\footnotesize{Observed SEDs of debris disks with submillimeter and millimeter data. The left panels are the 
photosphere-subtracted fluxes of the excess emissions with the best fitting slopes, while the right panels are the 
68\%, 95\% and 99\% confidence contours of the individual fits. The error contours also show the slope given by 
the \cite{dohnanyi69} mass distribution function in red, the value predicted by our numerical code in the green 
band, and the best global fit with errors in the blue band.}}
\label{fig:stars}
\end{center}
\end{figure*}

\addtocounter{figure}{-1}
\begin{figure*}[!t]
\begin{center}
\includegraphics[angle=0,scale=0.76]{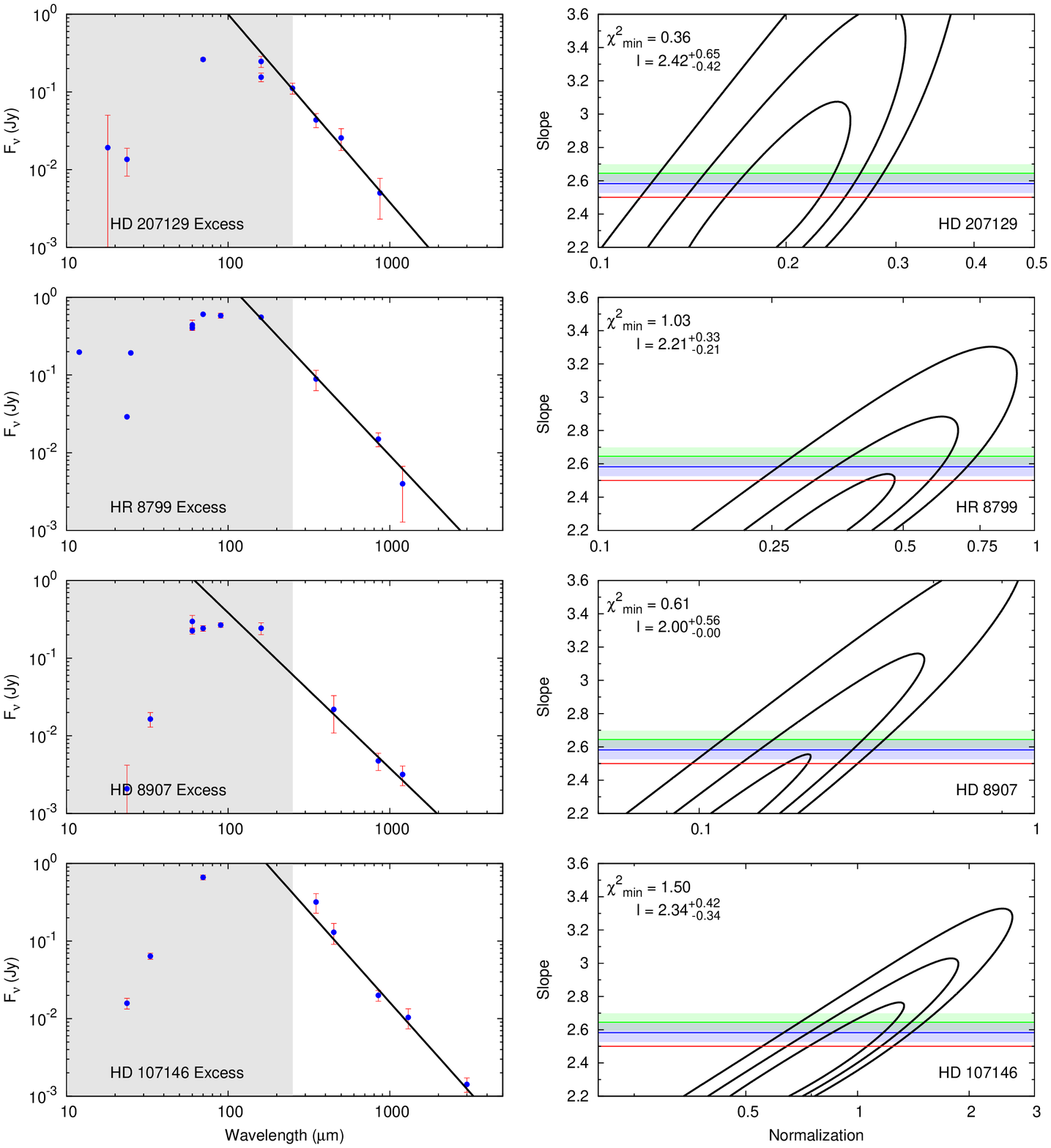}
\end{center}
\vspace{-2mm}{\bf Figure \ref{fig:stars}.\ (Cont.)}
\end{figure*}

We perform individual power-law fits to the data of each source as well as a fit to all sources simultaneously with 
a common Rayleigh-Jeans slope. In Figure \ref{fig:stars}, we present the photosphere subtracted excess emissions 
for each source in the left panels and plot the best-fit power-law spectrum of the form
\begin{equation}
F_{\nu} = {\rm A} \times \left(\frac{\lambda}{200~\micron}\right)^{-l}\;,
\end{equation}
obtained from individual fits. We show in the right panels of Figure \ref{fig:stars} the error contours of the slope
and normalization of the power-law at the 1, 2, and $3\sigma$ levels. The plots also indicate the 
$\chi^2_{\rm min}/{\rm d.o.f.}$ (the minimum of the reduced $\chi^2$) of each fit. The solid red line represents 
the Rayleigh-Jeans slope calculated from the \cite{dohnanyi69} analytic solution, the green band represents the 
best slope given by our reference model calculation (including errors from 50\% variations in the slope of the 
strength curve, see Figure \ref{fig:errs}), and the blue band yields our global fit solution of 
\begin{equation}
l=2.58\pm0.06\;.
\end{equation}
The global fit and our reference model agree within the errors of the prediction.

\section{Conclusions}

In this paper, we used our numerical model introduced in Paper I to follow the evolution of a distribution of particle 
masses. Our numerical model has been built 
to ensure mass conservation and that the resulting distribution of particles is not artificially offset due to numerical 
errors, as the integrations of the model span over 40 orders of magnitude in mass. In \S \ref{sec:var} of this paper, 
we demonstrate that lower precision integrations can lead to shallower particle distributions.

We varied all twelve collisional, all six system, and all three numerical variables of our model and examined 
the effects of these variations on the evolution of the particle mass distribution. 
The quasi steady-state particle distribution of the collisional system is extremely robust against variations in its variables, 
with the strongest effects occurring from changes to the tensile strength curve \citep{holsapple02,benz99}. Even 
these variations have mild effects on the slope of the particle mass distribution, modifying it only between the values 
of 1.84 and 1.94 (3.52 and 3.82 in size space, respectively). We find the dust mass distribution of our reference model to 
be $1.88$ (3.65 in size space). We find that waves occur when the collisional velocities are high or when particle 
strengths are low at the mass distribution cut-off, where the radiation force blowout dominates the dynamics.
Nonetheless, in \S \ref{sec:approx}, we show that a simple power-law mass (size) distribution is an appropriate approximation
even for extended disks with components only outside of 10 - $15~{\rm AU}$ for solar- and 50 - $60~{\rm AU}$ for early-type stars.

The Rayleigh-Jeans tail of the debris disk SEDs is dominated by the medium sized particles, whose mass distribution 
is less affected by possible wavy structures. We derive a simple formula that gives the slope of the 
measured flux density in the Rayleigh-Jeans part of the SEDs as
\begin{equation}
F_{\nu} \propto \lambda^{1-\eta_a} \nonumber\;.
\end{equation}
This implies that the mass distribution slope itself could, in principle, be measured from long-wavelength observations. 
We assemble a list of nine debris disks that have been measured at the far-IR, submillimeter, and millimeter wavelengths 
and examine the Rayleigh-Jeans slope of their emissions. Our predictions of a slope of $l = 2.65 \pm 0.05$ agrees well 
with the observations, which have a global slope fit of $l=2.58\pm0.06$.

\acknowledgments

Support for this work was provided by NASA through Contract Number 1255094 issued by JPL/Caltech. 
We thank Viktor Zubko and Karl Misselt for providing the numerical code to calculate the optical properties
of large grains.

\end{document}